\documentclass[reprint,superscriptaddress]{revtex4-1}

\usepackage{graphicx}
\usepackage{dcolumn}
\usepackage{bm}

\begin{document}

\newcommand\Tw{T_\mathrm{w}}
\newcommand\Teff{T_\mathrm{eff}}
\newcommand\Tf{T_\mathrm{f}}

\title{Electron Plasmas Cooled by Cyclotron-Cavity Resonance}

\author{A. P. Povilus}
\affiliation{Lawrence Livermore National Laboratory, Livermore, California, 94550 USA}
\affiliation{Department of Physics, University of California, Berkeley, California, 94720 USA }
\author{N.~D.~DeTal}
\affiliation{Department of Physics and Astronomy, Beloit College, Beloit, Wisconsin, 53511 USA }
\author{L. T. Evans}
\affiliation{Department of Physics, University of California, Berkeley, California, 94720 USA }
\author{N. Evetts}
\affiliation{Department of Physics and Astronomy, University of British Columbia, Vancouver, British Columbia, V6T~1Z4, Canada}
\author{J. Fajans}
\affiliation{Department of Physics, University of California, Berkeley, California, 94720 USA }
\author{W. N. Hardy}%
\affiliation{Department of Physics and Astronomy, University of British Columbia, Vancouver, British Columbia, V6T~1Z4, Canada}
\author{E. D. Hunter}%
\affiliation{Department of Physics, University of California, Berkeley, California, 94720 USA }
\author{I. Martens}%
\affiliation{Department of Chemistry, University of British Columbia, Vancouver, British Columbia, V6T~1Z4, Canada}
\author{F. Robicheaux}
\affiliation{Department of Physics and Astronomy, Purdue University, West Lafayette, Indiana, 47907 USA }
\author{S. Shanman}%
\author{C. So}%
\affiliation{Department of Physics, University of California, Berkeley, California, 94720 USA }
\author{X. Wang}
\affiliation{Department of Physics and Astronomy, Purdue University, West Lafayette, Indiana, 47907 USA }
\author{J. S. Wurtele}%
\affiliation{Department of Physics, University of California, Berkeley, California, 94720 USA }

\date{\today}

\begin{abstract}
We observe that high-$Q$ electromagnetic cavity resonances increase the cyclotron cooling rate of pure electron plasmas held in a Penning-Malmberg trap when the electron cyclotron frequency, controlled by tuning the magnetic field, matches the frequency of standing wave modes in the cavity.  For certain modes and trapping configurations, this can increase the cooling rate by factors of ten or more.  In this paper, we investigate the variation of the cooling rate and equilibrium plasma temperatures over a wide range of parameters, including the plasma density, plasma position, electron number, and magnetic field.

\end{abstract}

\maketitle

Cold, confined, non-neutral plasmas are complex, yet highly controllable physical systems that have a variety of potential applications, including basic plasma science \cite{davi:90,onei:16}, the production of monoenergetic beams for spectroscopic and material analysis \cite{dani:15}, and experiments studying the properties of antihydrogen \cite{amor:02,gabr:02,andr:10a}. These plasmas are typically confined in Penning-Malmberg traps, \cite{malm:75}, in which a homogeneous axial magnetic field restricts transverse motion, and electrostatic potentials, generated by a series of cylindrically symmetric electrodes, confine the axial motion. The magnetic field has the coincidental benefit that it causes the confined charged particles to execute circular, cyclotron orbits, thereby radiating away transverse energy \cite{onei:80a}. At sufficiently high magnetic fields, the cyclotron emission rate becomes fast enough for this mechanism to cool confined lepton plasmas \cite{malm:84,malm:88}; the axial degree of freedom \cite{hyat:87,beck:92} and additional trapped species \cite{gabr:89} may be sympathetically cooled through collisions.

The cyclotron emission rate depends on the density of electromagnetic field states which can absorb energy from the oscillating charges. In describing early NMR experiments, Purcell \cite{purc:46} argued that a single oscillator coupled to a resonant circuit sees an enhanced emission rate $\Gamma$ over the free-space rate $\Gamma_0$,
\begin{equation}
\frac{\Gamma}{\Gamma_0}=\frac{3Q \lambda^{3}} {4\pi^{2} V}.
\label{purcell}
\end{equation}
Here $\lambda$ is the wavelength of the radiation, $V$ is the volume of the resonator, and $Q$ is the quality factor.  For reference, the free space lepton cyclotron cooling rate at the cyclotron frequency $\omega_c$ is $\Gamma_0=(2/3)e^2\omega_c^2/3\pi\epsilon_0m_ec^3\approx0.26\, B^2[\mathrm{T}]\, \mathrm{s}^{-1}$ for leptons of charge $e$ and mass $m_e$. The factor of $2/3$ in this expression accounts for the collisional cooling of the axial degree of freedom from the two transverse degrees of freedom.

The Purcell effect has been studied in cold atoms \cite{kleppner:81}, semiconducting lasers \cite{yablo:87}, and cryogenic solid state systems \cite{bienfait:16}, but it has not previously been applied beyond the quasi single-particle regime to the cooling of non-neutral plasmas.

Penning-Malmberg traps often operate at fields of $\sim 1\,\mathrm{T}$.  The resulting cyclotron radiation wavelengths, $\lambda\sim 1\,\mathrm{cm}$, are comparable in size to the trap electrodes. With appropriate electrode geometries, the electrodes can trap high-$Q$ cavity modes. The resultant enhanced cyclotron coupling, and hence cooling, was first studied by Gabrielse and Dehmelt \cite{gabr:85} for single electrons, and later by Tan and Gabrielse \cite{tan:93} for relatively small clouds of non-equilibrium, parametrically-driven electrons.  In neither case were the resulting electron temperatures measured directly.  Here we study large electron clouds, indeed, electron plasmas, in thermal equilibrium, and present direct temperature measurements.

The single particle expression Eq.~(\ref{purcell}) does not give the correct cooling rate for non-neutral plasmas which can acquire an on-resonance impedance comparable to the vacuum cavity impedance set by $Q$. O'Neil \cite{onei:80a} suggested an optimization matching the cyclotron damping rate of the plasma to the (vacuum) linewidth of the cavity mode. Under these conditions (matched impedance), he calculated that the $N$-particle cooling rate has a maximum
\begin{equation}
\Gamma_{\mathrm{max}} =\sqrt{\frac{\pi e^2}{9 \epsilon_0 m_{\mathrm{e}}NV_{\mathrm{eff}}}}\approx 34\,\sqrt{\frac{\chi_{\rho}\,[\mathrm{m}^{-3}]}{N}} \,\, \mathrm{s}^{-1},
\label{gamma_max}
\end{equation}
where $\chi_{\rho}$ is the overlap integral which defines an effective inverse volume for each mode,
\begin{equation}
\chi_\rho = \frac{1}{V_{\mathrm{eff}}}= \frac {(1/N)\int dV \rho  E_{\perp}^{2}} {\int dV E^{2}},
\label{overlap}
\end{equation}
and $\rho$ is the plasma density. This factor takes into account the average field seen by the electrons. The plasmas in the experiments reported here are comparable in axial length scale to the cavity modes themselves, so we expect $\chi_\rho$ to depend on the plasma shape and position.

\begin{figure}
 \includegraphics[width=3in]{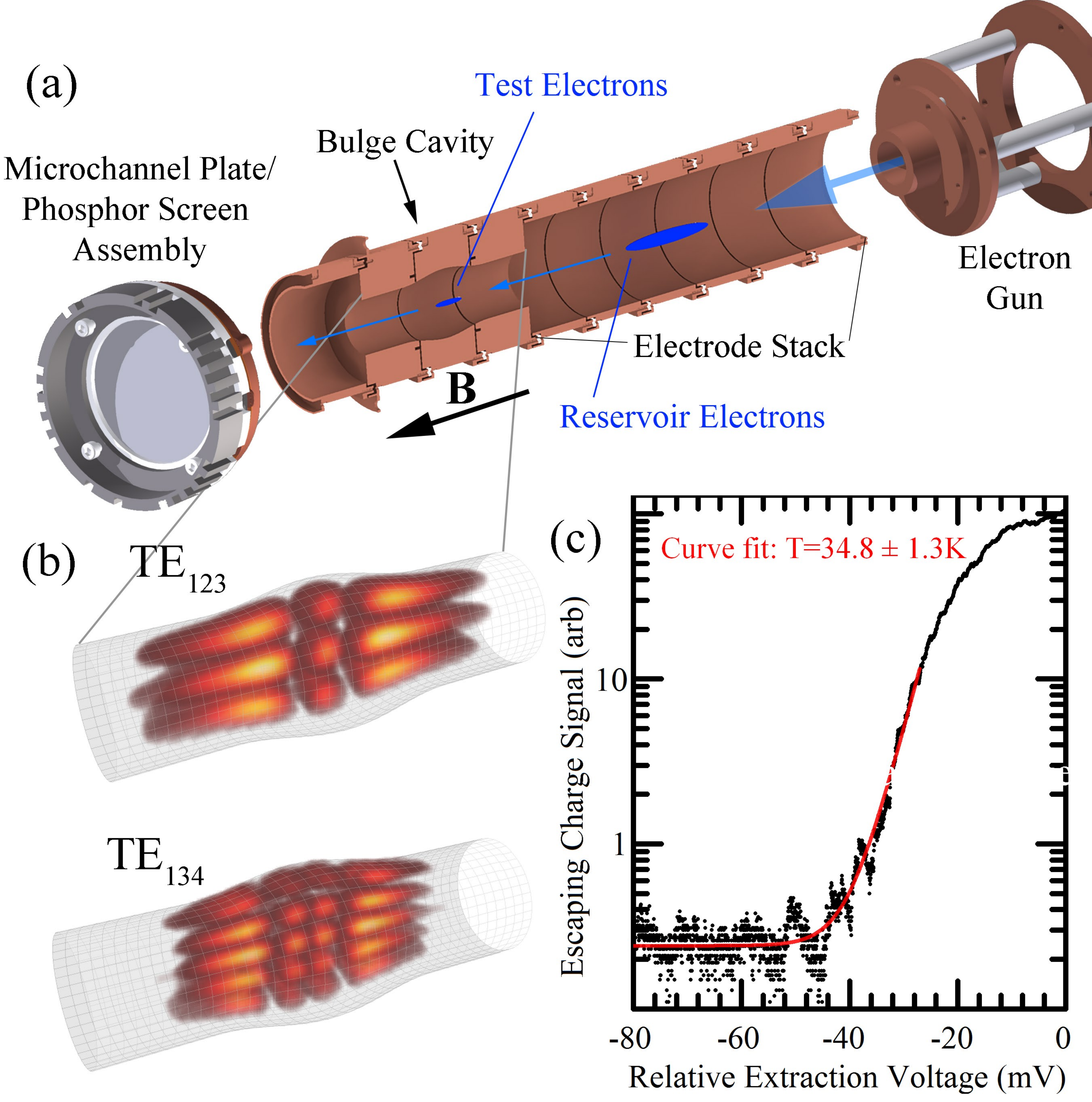}
 \caption{(a) Experiment schematic showing the electron gun, the microchannel plate, and the bulge cavity where modes are trapped. A ``reservoir'' of electrons ($L\approx 5\,\mathrm{cm}$, $r\approx 1\,\mathrm{mm}$, $N\approx 10^8\,\mathrm{e}^-$), and the test electrons ($L\approx 1\,\mathrm{cm}$, $r\approx 1\,\mathrm{mm}$, $N\approx 10^5\,\mathrm{e}^-$), are depicted in blue. (b) The $E_\perp$ cavity mode intensity patterns simulated by High-Frequency Structure Simulator \cite{HFSS}, shown for two cavity modes in cross section. Cavity mode indices can be interpreted as those of a right circular cylinder mode with the same topology. (c) Typical temperature measurement data (black) showing the escaped plasma charge as a function of the plasma confinement voltage, and the corresponding temperature fit (red). }
 \label{fig:schematic}
\end{figure}

Our experiments are done in a cryogenic electron plasma trap (Fig.~\ref{fig:schematic}). The trap is immersed in a strong axial magnetic field from a helium-cooled superconducting magnet, and electrons are generated by a thermionic emission electron gun. By manipulating the potentials on the $20\,\mathrm{mm}$ radius electrodes, we first trap a reservoir of $\sim 10^8$ electrons upstream, and then periodically transfer $10^3$--$10^6$ electrons downstream into the bulge cavity.  The electron transfer procedure reproducibly initializes the test electron cloud/plasma at a high temperature. The electrons then cool via cyclotron radiation, potentially with cavity enhancements.

The bulge cavity \cite{evet:15,evet:16} is formed from three electrodes ranging in radius from $10\,\mathrm{mm}$ to $12.5\,\mathrm{mm}$.  The cavity has a total length of about $38\,\mathrm{mm}$, and is open ended to allow for the transfer of electrons. The $Q$s of this cavity range from 300 to 2000, depending on the mode; here and below we report vacuum $Q$ values, noting that the presence of the plasma can reduce $Q$ and potentially scatter mode energy between the cavity and propagating waveguide modes. The cavity $Q$s were deliberately lowered to broaden the mode bandwidths by coating the cavity/electrode surfaces with nichrome.  The cavity surfaces are cooled to approximately $16 \,\mathrm{K}$.  In the absence of heating mechanisms, the electrons would come into thermal equilibrium with the effective temperature set by the combined effect of the cavity surfaces and the black body radiation that leaks in from the cavity ends; these ends are exposed to distant surfaces at higher temperatures. We would expect the cooling behavior to be dominated by these sources when not tuned to a cavity resonance.

We measure the plasma temperatures by raising one of the axial confining potentials $V(t)$ towards zero, thereby gradually releasing the plasma electrons.  The charge thus extracted is determined by first amplifying the plasma electron signal on the microchannel plate (MCP) [Fig.~\ref{fig:schematic}(a)], then converting the amplified signal to light on the adjacent phosphor screen, and  finally detecting the light [Fig.~\ref{fig:schematic}(c)] with a photomultiplier (not shown).  If, as we assume, the plasma is Maxwellian distributed, the charge released is initially proportional to $\exp(eV(t)/kT)$.  By fitting this curve (with a constant drift offset) to the data, we can obtain $T$, the plasma temperature \cite{eggl:92}.

We can repeat the full experimental cycle (transfer, relax and cool, release and measure $T$) about 100 times over the course of 5 minutes while we sweep the magnetic field or vary the parameters of the test plasmas. The plasma length $L$ and position $z$ are varied by changing the axial confining potentials, while the number $N$ is varied by adjusting the potentials used to transfer electrons from the reservoir. The magnetic field can be swept from $0$ up to $1.5\,\mathrm{T}$ with $\Delta B<0.03\, \mathrm{mT}$ precision.

\begin{figure}
 \includegraphics[width=3in]{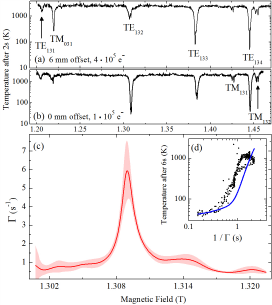}
 \caption{Field scan showing the temperature after 2 seconds of cooling for $10^5$ electrons held at (a) 6 mm and (b) 0 mm axial offset from trap center. The modes were identified by matching the cyclotron frequencies at the troughs to the bench-measured cavity resonances \cite{evet:15,evet:16}. (c) Cooling rate at TE$_{132}$ for a centered, long plasma. The bands represent 1$\sigma$ uncertainty on cooling rates derived from fits to the plasma temperature reached after 0.25, 0.38, 0.5, 0.75, 3, and 6 seconds of cooling at each field value. (d) Relationship between the cooling rate and the temperature reached after cooling for $6\,\mathrm{s}$ (black points). The blue curve is a solution to the cooling law Eq.~(\ref{cooling}) with the parameters $H=50\,\mathrm{K}\,\mathrm{s}^{-1}$, $T_i=23000\,\mathrm{K}$, $\Teff=35\,\mathrm{K}$.
 \label{fig:modes}}
\end{figure}

Figure~\ref{fig:modes}(a,b) shows the temperatures of plasmas held at two different axial locations. The plasmas were allowed to cool for $2\,\mathrm{s}$ while the magnetic field was swept at $0.02\,\mathrm{mT}\,\mathrm{s}^{-1}$. For $N > 10^5$ electrons, higher cooling rates (lower temperatures) were sometimes obtained when the overlap $\chi_\rho$ was relatively small. This result seems to disagree with Eqs.~(\ref{purcell}) and (\ref{gamma_max}). In particular, the TM$_{031}$ has E$_\perp$ = 0 on axis, so contributions to the overlap $\chi_\rho$ appear only for plasma electrons at a finite radius. This leads to a small $\chi_\rho \approx 0.03 \,\mathrm{cm}^{-3}$ for the plasmas in Fig.~\ref{fig:modes}(a). Yet this mode exhibits greater cooling power than the TE$_{131}$ and TE$_{132}$, for which $\chi_\rho \approx 0.5$ and $0.8 \,\mathrm{cm}^{-3}$ for the plasmas in Fig.~\ref{fig:modes}(a).

For a long plasma ($L=12\,\mathrm{mm}$) with $N\approx 2\times 10^5$ electrons centered on the TE$_{132}$ field minimum at $z=0$, we obtain an approximate Lorentzian line shape for the cavity enhancement [Fig.~\ref{fig:modes}(c)].  The maximum absolute cooling rate is $\Gamma \approx 6\,\mathrm{s}^{-1}$, a factor of $14$ enhancement over the free space emission rate for this mode. The fits and uncertainties were obtained via a standard bootstrap approach employing the \textsc{nls} and \textsc{ns} functions in \textsc{R} \cite{r:15}. The ability to independently vary the cooling rate allows us to obtain a relationship between $\Gamma$ and $T$. We compare our results to the differential cooling law with a heat source,
\begin{equation}
\frac{d\,T}{dt}=-\Gamma(T-\Teff)+H,
\label{cooling}
\end{equation}
where $H$ represents a constant background heating rate due to plasma expansion and radiofrequency noise coupling in through the electrode wires, and $\Teff$ is the effective temperature of the EM fields seen by the electrons, which is bounded from below by the electrode temperature ($16\,\mathrm{K}$), but may be raised by radiation from the electron gun, the MCP, and the plasma itself. To fit the data to Eq.~(\ref{cooling}) we assume an initial temperature $T_i=23200\,\mathrm{K}$ ($2 \,\mathrm{eV}$), but let the parameters $H$ and $\Teff$ be determined by the best fit to the longest cooling time data, shown in Fig.~\ref{fig:modes}(d). Although Eq.~(\ref{cooling}) ignores the time and temperature dependence of $\Gamma$ and $H$, a reasonable fit is obtained for higher cooling rate data. For $1/\Gamma\lesssim 0.5\,\mathrm{s}$, the plasma has already reached its final temperature $\Tf$ after $6\,\mathrm{s}$ and the data points all fall along the line $\Tf=\Teff+H/\Gamma$.

\begin{figure*}
 \includegraphics[width=6in]{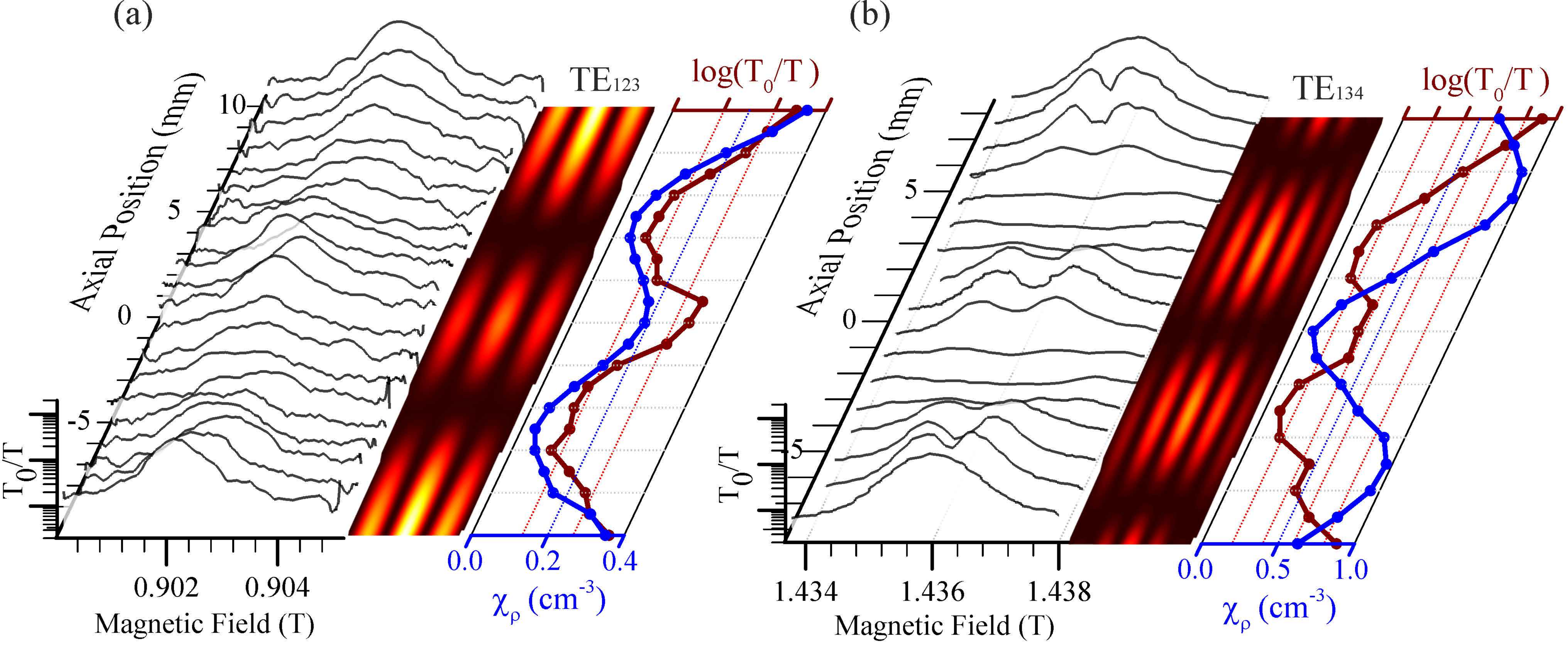}
 \caption{Cooling enhancements: (a) TE$_{123}$ with $2 \times 10^4$ electrons, (b) TE$_{134}$ with $3 \times 10^5$ electrons.  For each mode, the left waterfall plot shows the ratio $T_0/T$ as a function of the magnetic field and plasma position ($T_0$ is the typical off-resonant temperature for each dataset). The central color contour plot shows the mode structure.  The right graph plots $T_0/T$ at zero detuning [i.e., $0.902\,\mathrm{T}$ (a) and $1.436\,\mathrm{T}$ (b)] (red points), along with the overlap integral $\chi_\rho$ defined by Eq.~(\ref{overlap}) (blue points).}
 \label{fig:PosScan}
\end{figure*}

In Fig.~\ref{fig:PosScan} we plot the temperature for plasmas held for cooling at different axial positions and continuously varied magnetic field. The position and shape of a non-neutral plasma can be controlled to sub-mm precision, and the overlap integral Eq.~(\ref{overlap}) calculated using a zero-temperature solver \cite{so:14}, which combines the axial potential grid with the radial plasma density profile obtained by imaging the plasma when it is dumped onto the phosphor screen. Although the low-$N$ enhancements at TE$_{123}$ seem to match our expectation that higher coupling should lead to faster cooling, the high-$N$ data at TE$_{134}$ displays an unexpected pattern; at low electric fields (close to a node), the plasma was observed to have higher cooling rates.

\begin{figure}
 \includegraphics[width=3in]{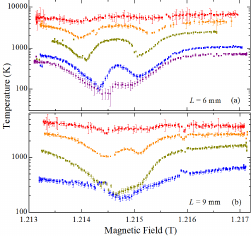}
 \caption{Lineshape splitting at TM$_{031}$ as a function of plasma length $L$ and temperature $T$. The confining potentials were set to produce 6 mm plasmas in (a) and 9 mm plasmas in (b). Plasma temperatures were measured after cooling for 0.1 (red), 0.5 (orange), 1.5 (green), 4 (blue), and 6 (purple) seconds. The feature visible at $1.2159\,\mathrm{T}$ is from a reservoir reload.}
 \label{fig:Splitting}
\end{figure}

Cyclotron lineshape splitting was previously observed by Tan and Gabrielse \cite{tan:93}, and described as being due to the modulation of the cavity field at the electron axial bounce
frequency. This effect should be especially large near a node of a cavity mode because the electric field seen by a bouncing electron goes to zero as it passes through the node. Since, in many cases, the collision rate in the plasma is less than the frequency of axial oscillation, we can clearly observe this effect in Figs.~\ref{fig:PosScan}(b) and~\ref{fig:Splitting}. The splitting can be manipulated by changing the plasma parameters. For example, by changing the electrode potentials, we can go from a more strongly confining potential [$L=\,6\,\mathrm{mm}$ in Fig.~\ref{fig:Splitting}(a)] with a larger frequency to a less strongly confining potential [$L=\,9\,\mathrm{mm}$ in Fig.~\ref{fig:Splitting}(b)] with a lower frequency. It is clear in Fig.~\ref{fig:Splitting} that the splitting for $L=\,6\,\mathrm{mm}$ is much larger than for $L=\,9\,\mathrm{mm}$; preliminary calculations of the frequency splitting match this observation. The splitting also tends to decrease as the temperature decreases for fixed electrode potentials. This effect is most clear in Fig.~\ref{fig:Splitting}(a).  At lower temperature, the Debye length becomes shorter and the plasma flattens the potential well, resulting in a lower bounce frequency and less splitting.

\begin{figure}
 \includegraphics[width=3in]{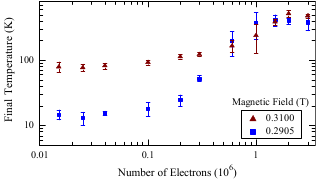}
 \caption{Equilibrium temperature of plasmas with $10^4 \text{--} 10^6$ electrons. In the dark red dataset the field is detuned $19.5\,\mathrm{mT}$ ($546\,\mathrm{MHz}$) from the TE$_{111}$ resonance.}
 \label{fig:NumScan}
\end{figure}

Purcell's formula Eq.~(\ref{purcell}) shows that, unlike free-space cyclotron emission, for which $\mathrm{\Gamma_0 \propto B^2}$,  cavity-enhanced emission should permit fast cooling at relatively low magnetic fields. This effect can be demonstrated by going to a field of 0.29 T and letting the plasmas cool to their minimum temperature. The dominant cooling mechanism at such low magnetic fields is normally thought to be collisions on the background gas;  since we held these plasmas for $36\,\mathrm{s}$, non-cavity cyclotron cooling ($1/\Gamma_0 \approx 46\,\mathrm{s}$ at 0.29 T) should have played only a limited role. But because of the TE$_{111}$  resonance at 0.2905 T, we obtain a dramatic reduction of the minimum plasma temperature simply by tuning the magnetic field from 0.31 to 0.29 T. Although the $Q$ for this mode is low ($Q \approx 300$), the resonant cooling reduces the lowest achievable temperature for small numbers of electrons by nearly an order of magnitude (Fig.~\ref{fig:NumScan}).

For larger numbers of electrons there is little benefit to operation at the TE$_{111}$ resonance, as indicated by the convergence of the on- and off-resonant curves in Fig.~\ref{fig:NumScan}. Since the cavity mode cannot be at a higher temperature than the plasma and still cool the plasma, there is an upper bound $\frac{1}{2}kT\cdot \omega_\lambda/Q$ for the rate at which energy can be removed from the system, assuming a single resonant mode with frequency $\omega_\lambda$ interacting with an $N$-electron plasma at temperature $T$. This leads to the bottleneck condition $N \Gamma \sim \omega_\lambda / Q$ \cite{povi:15}.  For the TE$_{111}$ this approximation gives $N \sim 10^7$ electrons, so this bottleneck argument does not explain the temperature increase in  Fig.~\ref{fig:NumScan}, which begins at $N \sim 10^5$ electrons. Approximately the same limiting $N$ was observed in cooling data taken at the TE$_{121}$ ($Q \approx 1800$) and TE$_{131}$ ($Q \approx 2600$).

A different bottleneck occurs when nearly identical oscillating dipoles strongly couple to a cavity; such systems can be decomposed into superradiant and subradiant modes. For simplicity, consider a case where only one superradiant mode is dominant. This mode has a decay rate $N$ times the single-particle rate $\Gamma_1$, while all other modes have much lower decay rates. Only a small fraction of the total system energy is ever in the dominant superradiant mode; once this energy damps away, cooling slows dramatically if the mode is not repopulated. However, dephasing can continuously repopulate the mode. Such dephasing might be caused, for instance, by small variations in the cyclotron frequency $\Delta\omega_c$ across the plasma. If $\Delta \omega_c$ is greater than $N\Gamma_1$, approximate equipartition will be maintained, and the plasma will continue to cool with rate $\Gamma_1$. For our TE$_{111}$ mode, we estimate $\Gamma_1 \sim 10 \,\mathrm{s}^{-1}$ and $\Delta \omega_c \sim 2\pi\cdot 5 \,\mathrm{MHz}$. Thus, quasi-equipartition will be maintained for $N \lesssim 5 \times 10^5$ electrons, and we would expect cooling to slow for larger $N$; this is close to the transition observed in Fig.~\ref{fig:NumScan}.  Even when $N$ is below this bound, the cooling rate $\Gamma_{\mathrm{max}}$ predicted by Eq.~(\ref{gamma_max}) is only obtained  \cite{onei:80a}  if $\Delta \omega_c$ is tuned to match the cavity linewidth. We cannot directly control $\Delta \omega_c$, and it may evolve as the plasma cools, so it is not surprising that even the largest rate observed in our experiment, $6\, \mathrm{s}^{-1}$ [for the TE$_{132}$ cavity mode, see Fig.~\ref{fig:modes}(c)] was less than the corresponding predicted rate, $\Gamma_{\mathrm{max}}\approx 30 \, \mathrm{s}^{-1}$.

In conclusion, we have demonstrated cyclotron-cavity resonant cooling of pure electron plasmas with large numbers of electrons. We implemented this technique in an open-ended geometry compatible with standard Penning-Malmberg trap experiments, as well as with experiments for performing antimatter spectroscopy and molecular spectroscopy using positrons. The cooling rate was found to be influenced by a wide range of plasma and trap parameters, including the mode profile and the plasma density, length, and temperature. By optimizing the plasma-cavity overlap, cooling enhancements of up to 14 were obtained with $N > 10^5$ electrons. For these large $N$ plasmas, an unexpected but essential requirement for optimal cooling is that the plasma must be located far from the field maximum.

The fact that, under certain circumstances, better cooling appears to be obtained when the plasma is close to the field null has never, to our knowledge, been observed before. This striking behavior has been observed in our experiment for all accessible modes having nodes at the cavity center (TE$_{122}$, TE$_{132}$, TE$_{134}$) as well as those with antinodes at the center, which require placing the plasma at a $4 \text{--} 5\,\mathrm{mm}$ axial offset (TE$_{123}$, TE$_{133}$). Preliminary data suggests that cooling is still enhanced at the nodes for $N \sim 10^6$ electrons. Therefore cavity enhanced cooling remains a very attractive possibility for antimatter research, which requires cold plasmas containing millions of electrons and positrons. Future experiments will pursue investigations of plasma cavity cooling, particularly at high $N$, and further explore the node/antinode optimization.

\begin{acknowledgments}
We thank M. Baquero-Ruiz, S. Chapman, M. Turner, and N. Lewis for their help during the building and testing stages of the experiment. This work was supported by the DOE DE-FG02-06ER54904 and DE-SC0014446, the NSF 1500538-PHY and 1500470-PHY, the LLNL DE-AC52-07NA27344, and the NSERC SAPPJ-2014-0026.
\end{acknowledgments}



\end{document}